\documentclass[journal]{IEEEtran}
\usepackage{cite}
\usepackage{graphicx}
\usepackage{caption}
\usepackage{pbox}
\usepackage{rotating}
\usepackage{placeins}
\graphicspath{ {graphics/} }
\usepackage[caption=false,font=footnotesize]{subfig}
\usepackage[algo2e]{algorithm2e}
\usepackage{algpseudocode}
\usepackage{array}
\usepackage{url}
\usepackage{amsmath}
\usepackage{amssymb}
\usepackage{amsfonts}
\usepackage{epstopdf}
\usepackage{xcolor}

\SetAlCapSkip{1em}

\def \R {{\mathbf R}}
\def \M {{\mathbf M}}
\def \N {{\mathbf N}}
\def \T {{\mathbf T}}
\def \x {{\mathbf x}}
\def \y {{\mathbf y}}

\def \v {{\mathbf v}}
\def \u {{\mathbf u}}
\def \s {{\mathbf s}}
\def \m {{\mathbf m}}

\def \veczero {{\mathbf{0}}}
\def \xopt{{\x^\ast}}

\def \ssMN {{\s^2\M\N}}

\def \MN {{\M\N}}

\def \HRimspace {{\R^{\s\M \times \s\N}}}
\def \HRspace {{\R^{\s\M \times \s\N \times \T}}}

\def \LRimspace {{\R^{\M \times \N}}}
\def \LRspace {{\R^{\M \times \N \times \T}}}

\def \BlurImSpace {{\R^{\ssMN \times \ssMN}}}

\def \ScaleImSpace {{\R^{\MN \times \ssMN}}}

\def \ScaleMat  {{\mathbf S}}
\def \BlurMat {{\mathbf H}}
\def \FlowMat {{\mathbf F}}
\def \DegradeMat {{\mathbf G}}

\def \sqnorm[#1]{{{ \Arrowvert #1  \Arrowvert}}^2_2}
\DeclareMathOperator*{\argmin}{arg\,min}

\def \PrimGap {{\sqnorm[\ScaleMat\BlurMat\x - \y]}}

\ifCLASSINFOpdf
\else
\fi
\begin{document}
\title{Unified Single-Image and Video Super-Resolution \\ via Denoising Algorithms}

\author{Alon~Brifman,
        Yaniv~Romano,
        and~Michael~Elad,~\IEEEmembership{Fellow,~IEEE}% <-this % stops a space
}

\maketitle

\begin{abstract}
Single Image Super-Resolution (SISR) aims to recover a high-resolution image from a given low-resolution version of it.
Video Super Resolution (VSR) targets series of given images, aiming to fuse them to create a higher resolution outcome.
Although SISR and VSR seem to have a lot in common,
most SISR algorithms do not have a simple and direct extension to VSR.
VSR is considered a more challenging inverse  problem, mainly due to its reliance on a sub-pixel accurate motion-estimation, which has no parallel in SISR.
Another complication is the dynamics of the video, often addressed by simply generating a single frame instead of a complete output sequence.  

In this work we suggest a simple and robust super-resolution framework that can be applied to single images and easily extended to video.
Our work relies on the observation that \textit{\textbf{denoising}} of images and videos is well-managed and very effectively treated by a variety of methods. 
We exploit the Plug-and-Play-Prior framework and the Regularization-by-Denoising (RED) approach that extends it, and show how to use such denoisers in order to handle the SISR and the VSR problems using a unified formulation and framework.
This way, we benefit from the effectiveness and efficiency of existing image/video denoising algorithms, while solving much more challenging problems.
More specifically, harnessing the VBM3D video denoiser, we obtain a strongly competitive motion-estimation free VSR algorithm, showing tendency to a high-quality output and fast processing.
\end{abstract}

\begin{IEEEkeywords}
Single Image Super-Resolution, Video Super-Resolution,  Plug-and-Play-Prior, RED, Denoising, ADMM
\end{IEEEkeywords}

% \fi
\IEEEpeerreviewmaketitle

\section{Introduction}\label{SS:intro}

The single-image super-resolution (SISR) problem assumes that a given measured image $\y$ is a blurred, spatially decimated, and noisy version of a high quality image $\x$. Our goal in SISR is the recovery of $\x$ from $\y$. This is a highly ill-posed inverse problem, typically handled by the Maximum a'posteriori Probability (MAP) estimator. Such a MAP strategy relies on the introduction of an image prior, representing the minus log of the probability density function of images. Indeed, most of the existing algorithms for SISR differ in the prior they use -- see for example \cite{SISR_ADAPTIVE_SPARSE, NCSR_REF, SISR_VIA_SPARSE_REPR, SISR_STAT_SPARSE,SISR_TV,SISR_PATCH_REDUNDENCY, SISR_SELF_SIMILARITY, SISR_SELF_SIMILARITY_MULTISCALE}. We should mention that convolutional neural networks (CNN) have been brought recently as well to serve the SISR \cite{SISR_DEEP_CNN, SISR_DEEP_NETWORK, IRCNN_REF}, often times leading to state-of-the-art results.

The Video Super Resolution (VSR) task is very similar to SISR but adds another important complication -- the temporal domain. Each frame in the video sequence is assumed to be a blurred, decimated, and noisy version of a higher-resolution original frame. Our goal remains the same: recovery of the higher resolution video sequence from its measured degraded version. However, while this sounds quite close in spirit to the SISR problem, the two are very much different due to the involvement of the temporal domain. Indeed, one might be tempted to handle the VSR problem as a simple sequence of SISR tasks, scaling-up each frame independently. However, this is highly sub-optimal, due to the lack of use of cross relations between adjacent frames in the reconstruction process.

More specifically, the VSR task can be formulated using the MAP estimator in a way that is similar to the formulation of the SISR problem. Such an energy function should include a log-likelihood term that describes the connection between the desired video and the measured one, and a video prior. While the first expression is expected to look the same for SISR and VSR, the video prior is likely to be markedly different, as it should take into account both the spatial considerations, as in the single image case, and add a proper reference to the temporal relations between frames. Thus, although SISR and VSR have a lot in common, the suggested priors for each task differ substantially, and hence SISR algorithms do not tend to have an easy adaptation to VSR.

The gap between the two super-resolution problems explains why VSR algorithms tend to use entirely different methods to tackle their recovery problem, rather than just extending SISR methods. Classic VSR methods commonly turn to explicit subpixel motion-estimation\footnote{Even CNN based solutions have been relying on such motion estimation and compensation.} \cite{WARP_REF, FAST_ROBUST_MFSR, SUB_PIXEL_REF2, SR_OPTICAL_FLOW, SUBPIXEL_SHIFT, PURE_TRANS, SR_IMAGE_SEQUENCE, ROBUST_SR, DUAL_OPTICAL_FLOW, MOTION_ESTIMATION_REF1, DEEPSR_REF, VIDEOSR_CNN, SUBPIXEL_CNN}, which has no parallel in SISR. For years it was believed that this ingredient is unavoidable, as fusing the video frames amounts to merging their grids, a fact that implies a need for an accurate sub-pixel registration between these images. Exceptions to the above are the two algorithms reported in \cite{GENERAL_NLM, 3DSKR_REF}, which use implicit motion-estimation, and thus are capable of handling more complex video content.

This work's objective is to create a robust joint framework for super resolution that can be applied to SISR and easily be adapted to solve the VSR problem just as well. Our goal is to formulate the two problems in a unified way, and derive a single algorithm that can serve them both. The key to the formation of such a bridge is the use of the Plug-and-Play-Prior (PPP) method \cite{PPP_REF}, and the more recent framework of Regularization-by-Denoising (RED) \cite{RED_REF}. Both PPP and RED offer a path for turning any inverse problem into a chain of denoising steps. As such, our proposed solution for SISR and VSR would rely on the vast progress made in the past two decades in handling the image and video denoising problems.

This work presents a novel and highly effective SISR and VSR recovery algorithm that uses top-performing image and video denoisers. More importantly, however, this algorithm has the very same structure and format when handling either a single image or a video sequence, differing only in the deployed denoiser. In our previous work \cite{OUR_SISR_REF} we have successfully harnessed the PPP scheme to solve the SISR problem. In this paper our focus is on extending this formulation to VSR while keeping its architecture. With the recent introduction of the Regularization-by-Denoising (RED) framework \cite{RED_REF}, we use this as well in the migration from the single image case to video. As demonstrated in the results section, the proposed paradigm is not only simple and easy to implement, but also leads to state-of-the-art results in video super-resolution, favorably competing with the best available alternatives.

The rest of this paper is organized as follows: Section \ref{SS:background} introduces the PPP and RED schemes, which play a critical role in this work. Section \ref{SS:framework} presents our suggested framework and its properties, and section \ref{SS:Results} provides extensive experimental results. Section \ref{SS:Conclusion} concludes this work.

\section{Background on PPP and RED}\label{SS:background}
In this section we present the Plug-and-Play-Prior \cite{PPP_REF} and Regularization-by-Denoising \cite{RED_REF} schemes, which are central to our framework. This section mostly follows \cite{PPP_REF, RED_REF, ADMM_REF}.

\subsection{Plug-and-Play-Prior} \label{SS:PPP}
Many inverse problems (including the super-resolution ones) are formulated as a MAP estimation, a factored sum of two expressions, or penalties: a data fidelity term (usually the log-likelihood) and a prior function. Such a general inverse problem may appear as follows:
\begin{equation}
\label{MAP_cost}
\xopt = \argmin_\x \frac{1}{2}\sqnorm[\DegradeMat\x - \y] + \beta R\left(\x\right),
\end{equation}
where $\x$ is the unknown image to be recovered, $\y$ is the measured image, assumed to be a noisy contaminated version of $\DegradeMat \x$, $\DegradeMat$ being the degradation operator. The functional $R(\cdot)$ stands for the image prior, and the parameter $\beta$ multiplying it sets the relative weights between the two penalties.

The Plug-and-Play-Prior (PPP) scheme \cite{PPP_REF} offers a method to separate the two in a manner that allows us to use  prior functions that are already integrated into Gaussian denoising algorithms. Thus, we use the denoiser as a black-box tool, while solving for another, more challenging, inverse problem. Let us illustrate the PPP on the problem posed in Equation (\ref{MAP_cost}). Using variable splitting, we can separate the degradation model (the $ \ell_2 $ data fidelity term) from the prior:
\begin{align}
\label{MAP_cost_split}
\begin{split}
\xopt = & \argmin_\x \frac{1}{2}\sqnorm[\DegradeMat\x - \y] + \beta R(\v), \\
 &\text{s.t} \quad \x = \v.
\end{split}
\end{align}
Using the Augmented Lagrangian strategy, the constraint can be turned into an additive penalty,
\begin{align}
\label{Augmented_Lagrangian}
\xopt =  \argmin_\x \frac{1}{2}\sqnorm[\DegradeMat\x - \y] + \beta R(\v) + \frac{\rho}{2}\sqnorm[\x - \v + \u],
\end{align}
where $\u$ is the scaled Lagrange multipliers vector, and $\rho$ is a parameter to be set\footnote{Often times, it is found beneficial to increase this parameter throughput the iterative algorithm given below.}.
Applying the ADMM \cite{ADMM_REF}, we obtain the following iterative scheme to minimize Equation \eqref{MAP_cost_split}:
\begin{subequations}
\label{ADDM_iteration}
\begin{align}
\label{x_update}
\x^{k+1} & = \argmin_\x \frac{1}{2}\sqnorm[\DegradeMat\x - \y] + \frac{\rho}{2}\sqnorm[\x - \v^{k} + \u^k] \\
\label{v_update}
\v^{k+1} & = \argmin_\v \beta R(\v) + \frac{\rho}{2} \sqnorm[\x^{k+1} - \v+ \u^k] \\
\label{u_update}
\u^{k+1} & = \u^{k} + \x^{k+1} - \v^{k+1}
\end{align}
\end{subequations}
Notice that Equation (\ref{x_update}) is quadratic in $\x$ and therefore can be solved analytically. Moving to Equation (\ref{v_update}), this can be re-written as
\begin{align}
\label{v_update_rewriten}
\v^{k+1} & = \argmin_\v \beta R(\v) + \frac{1}{2(\frac{1}{\sqrt{\rho}})^2} \sqnorm[\v- \widetilde{\v}],
\end{align}
where $\widetilde{\v} = \x^{k+1}+\u^{k}$. The above is nothing but a denoising problem, aiming at cleaning the noisy image $\widetilde{\v}$, with $\sigma = \frac{1}{\sqrt{\rho}}$. Hence we can use the denoiser as a black-box tool for solving step (\ref{v_update}), even without having access to the explicit prior $R(\v)$ we are relying on.

\subsection{Regularization by Denoising}\label{SS:RED}
Like PPP, Regularization by Denoising (RED) is another scheme that integrates denoisers into a reconstruction algorithm in order to solve other inverse problems.
Yet, as opposed to the PPP scheme, RED defines an explicit prior, constructed by a chosen denoising algorithm. Then RED solves the MAP estimator with the defined prior in order to achieve a global optimizer (under mild conditions).

More specifically, given a denoising function $f(\x)$, RED sets the prior to be $R(\x) = \frac{1}{2}\x^T\left(\x - f\left(\x\right)\right)$.
This is an image-adaptive Laplacian-based regularization functional, penalizing over the inner product between a signal $ \x $ and its denoising residual $ \x-f(\x) $.
Under the assumptions that (i) $ \nabla f(\x) $ exists (i.e. $f$ is differentiable) and is symmetric,
(ii) $ f\left(c\x\right) = cf\left(\x\right)$, for $ c\rightarrow 1 $ (local homogeneity), and
(iii) the spectral radius of $ \nabla f(\x) $ is smaller or equal to 1, the authors of \cite{RED_REF} prove two key properties:

(i) The gradient of the prior expression is nothing but the denoising residual,
\begin{align*}
R(\x)=\frac{1}{2} \x^T(\x-f(\x)) \rightarrow \nabla R(\x) = \x-f(\x).
\end{align*}

(ii)  The convexity of the suggested prior (and thus the overall objective) is guaranteed, meaning that the MAP optimization process will yield a global minimizer.

It turns out that various state-of-the-art denoising algorithms satisfy these conditions \cite{RED_REF}, or nearly so, posing RED as an appealing alternative to the PPP approach.

Equation (\ref{MAP_cost}) with the newly introduced prior may be solved using several different strategies, as indeed was done in \cite{RED_REF}. One of the proposed strategies uses ADMM, derived in a similar manner to the one in Section \ref{SS:PPP}.
However, this time we have a specific regularizer at hand.
Concentrating on step (\ref{v_update}), by plugging the RED regularizer we get
\begin{align}
\label{RED_v_update}
\v_{k+1} & = \argmin_\v  \frac{\beta}{2}\v^T\left(\v - f\left(\v\right)\right) + \frac{\rho_{k}}{2} \sqnorm[\x_{k+1} - \v+ \u].
\end{align}
Exploiting the relation  $ \nabla R(\x) = \x - f(\x)$, and setting the gradient of this cost function to zero, we obtain
\begin{align}
\label{RED_v_update_grad_simplified}
\beta\left(\v - f\left(\v\right)\right) + \rho\left(\v - \x_{k+1} - \u\right) = 0,
\end{align}
which can be solved by the fixed point strategy, leading to the following update rule for $ \v $ (please refer to \cite{RED_REF} for more details):
\begin{align}
\label{RED_v_update_fixed_point}
\v_j = \frac{1}{\beta +  \rho}\left(\beta f\left(\v_{j-1}\right) +  \rho\left(\x_{k+1} + \u\right)\right).
\end{align}
The signal $ \v_j $ is the estimate of the $ j $-th step of the fixed point method that minimizes Equation \eqref{RED_v_update}.
As a result, once again, the solution of $\v$ amounts to the application of a denoiser as a black-box tool (possibly for several iterations).\footnote{We note that in \cite{RED_REF}, an alternative scheme to the ADMM was proposed based on a direct Fixed-Point strategy.}

In summary, we have in our hands powerful tools to take denoisers and use them in order to solve other, more involved, inverse problems. Our next step is to harness these to the super resolution problem, both for  single images and video.

\section{The Proposed Super-Resolution Framework}\label{SS:framework}

In this section we present our novel unified formulation of the SISR and the VSR problems, and the common algorithm that served them both. We start by describing the SISR problem, and then move to the video counterpart. Our next step is to describe our PPP/RED based approach to super-resolution reconstruction, and as this relies on denoising algorithms, we precede this by mentioning the existing state-of-the-art denoising methods for stills  and video. We conclude this section by discussing computational complexity and convergence issues.

\subsection{The Singe-Image Super-Resolution Problem}
The single-image super-resolution (SISR) problem starts with an unknown High-Resolution (HR) image $\x \in \HRimspace$ ($s>1$), of which we are only given a blurred, spatially
decimated (in a factor $ \s $ in each axis), and noisy Low-Resolution (LR) measurement $\y \in \LRimspace$.
Our aim is to recover $\x$ from $\y$.
This inverse problem can be formulated by the following expression that ties the measurements to the unknown:
\begin{align}
\label{IR_SR_eq}
\y = \ScaleMat\BlurMat\x + \eta.
\end{align}
The matrix $\BlurMat \in \BlurImSpace$ blurs the original image, $\ScaleMat \in \ScaleImSpace$ is the down-sampling operator and $\eta \sim N(0; \sigma^2 \mathbf{I}) \in \LRimspace $  is an additive zero-mean
white Gaussian noise. Note that $\x$, $\y$ and $\eta$ are all held as column vectors, after lexicographic ordering, that is, they are vectors of length
$\ssMN$ and $\MN$ respectively, where the columns of the image are concatenated one after another to form a long one-dimensional vector.

The Maximum Likelihood (ML) estimator for this problem is defined by
\begin{align}
\label{ML_SR_cost}
\xopt = \argmin_\x \PrimGap,
\end{align}
yet the ill-posed nature of this problem ($\ScaleMat\BlurMat$ is non-invertible) renders this approach useless. Using the MAP estimator instead leads us to minimize the ML, augmented with a predefined prior. That is, Equation (\ref{ML_SR_cost}) should be regularized,
\begin{align}
\label{MAP_SR_cost}
\xopt = \argmin_\x \PrimGap + R\left(\x\right),
\end{align}
$R(\x)$ being the prior, discriminating ``good looking'' images from ``bad'' ones by giving the ``good'' images a lower numerical value.
And so, the quest of the holy grail, so to speak, begins: What is the appropriate prior to use for images? Vast amount of work has been invested in addressing this question. Indeed, most of the existing algorithms for SISR differ in the prior they use, or the numerical method they deploy for solving the presented optimization problem (\ref{MAP_SR_cost}).
Commonly chosen priors are based on sparse representations \cite{SISR_ADAPTIVE_SPARSE, NCSR_REF, SISR_VIA_SPARSE_REPR, SISR_STAT_SPARSE}, spatial smoothness such as Total Variation \cite{SISR_TV}, self-similarity \cite{SISR_PATCH_REDUNDENCY, SISR_SELF_SIMILARITY, SISR_SELF_SIMILARITY_MULTISCALE}, and more.

In recent years convolutional neural networks (CNN) have been used for SISR quite successfully \cite{SISR_DEEP_CNN, SISR_DEEP_NETWORK, IRCNN_REF}.
Observe that this approach bypasses the explicit use of the MAP formulation, replacing it by a direct learning of the recovery process from the input low-resolution image to the desired high-resolution output. One may argue that this supervised approach incorporates the MAP strategy and the prior in it implicitly, by shaping the solver as a minimizer of the MAP energy task.

\subsection{Moving to Video Super-Resolution}
The Video Super Resolution (VSR) task may seem to be similar to the SISR problem, but it adds another complicating factor -- the temporal domain. Here, the HR video, $\x \in \HRspace$, is composed of $\T$ frames that are assumed to have some relation between them, manifested as motion or flow. Often, as introduced in \cite{WARP_REF} and later used in \cite{DEEPSR_REF, FAST_ROBUST_MFSR, MOTION_ESTIMATION_REF1,SR_OPTICAL_FLOW}, a warp or motion operator is used to express this relation between the frames,
\begin{align}
\label{Flow_Notation}
\x_{i} = \FlowMat_{i}\x_{0} + \m_{i},
\end{align}
where $\x_{i}$ is the i-th frame, $\FlowMat_{i}$ is the motion operator that takes us from the 0-th frame to the i-th one, and $\m_{i}$ is new information appearing in $\x_{i}$.
The given low-resolution video $\y \in \LRspace$ is obtained from $\x$ via the same relation as in Equation (\ref{IR_SR_eq}), where the operators $\BlurMat$ and $\ScaleMat$ apply their degradations on the entire video sequence $\x$,  operating on each frame independently\footnote{In fact, a spatio-temporal blur can be easily accommodated as well by the algorithms discussed in this paper.}.

Using the MAP estimator for the VSR task can be formulated exactly as in Equation (\ref{MAP_SR_cost}), where $\x$ and $\y$ are now the complete high-resolution and low-resolution video sequences. Here again we face the need to find an appropriate prior that could grade video quality, and as already mentioned in the Introduction, such a prior is expected to be very different from a single image one, due to the need to refer to the temporal relations in Equation (\ref{Flow_Notation}).

And so, although some of the SISR algorithms mentioned above are considered state-of-the-art, and although SISR and VSR have very similar formulations, none of these known algorithms was adapted to VSR (apart from the trivial adaptation of performing SISR for each frame independently).
Put very simply, SISR algorithms do not generally have an easy adaptation to VSR.
Indeed, VSR algorithms tend to use entirely different methods to tackle their recovery problem, built around an explicit optical-flow or (subpixel-)motion-estimation process, which has no parallel in SISR. Hence, most of the VSR algorithms are heavily dependent on such highly performing motion-estimation algorithms, a fact that leads to more costly overall recovery processes \cite{WARP_REF, DEEPSR_REF, FAST_ROBUST_MFSR, MOTION_ESTIMATION_REF1, SR_OPTICAL_FLOW, SUB_PIXEL_REF2, VIDEOSR_CNN,SUBPIXEL_SHIFT, PURE_TRANS, SR_IMAGE_SEQUENCE, ROBUST_SR, SUBPIXEL_CNN, DUAL_OPTICAL_FLOW}.
An additional unfortunate by-product of this strategy is an extreme sensitivity of these methods to motion-estimation errors, causing sever artifacts in the resulting frames. As a consequence, classic VSR algorithms are known to be limited in their ability to process videos with only simple and global motion trajectories.

As a side note, we mention the following: The work reported in \cite{GENERAL_NLM} is the first VSR algorithm to abandon \emph{explicit} motion-estimation by generalizing NLM \cite{NLM_REF} for super-resolution. The 3DSKR algorithm\cite{3DSKR_REF} followed it, replacing the accurate motion-estimation by a multidimensional kernel regression.
Both these algorithms and their follow-up work \cite{3DSKR_FOLLOWUP} are capable of processing videos with far more complex motion. Still, these methods rely on the computation of weights based on every pixel's neighbourhood or patch, making them computationally heavy.
As in the SISR, CNNs have made their appearance to the VSR problem as well \cite{MOTION_ESTIMATION_REF1,  DEEPSR_REF, VIDEOSR_CNN, SUBPIXEL_CNN}, yet these methods too often integrate motion-estimation into their algorithmic process, hence being computationally heavy.

\subsection{Advancements in Denoising Algorithms}
We move now to discuss a simpler inverse problem -- denoising -- the removal of additive noise from images and video.
This task is a special case of the SISR and VSR problems, obtained by setting $\ScaleMat=\BlurMat={\mathbf I}$.
While this may appear as a diversion from this paper's main theme, the opposite is true. As we rely on the PPP or the RED schemes to construct an alternative super-resolution reconstruction algorithm, denoising algorithms are central to our work.

Image and video denoising have made great advancements over the past two decades, resulting in highly effective and efficient algorithms.
As denoising is the simplest inverse problem, such algorithms clearly encompass in them some sort of a prior knowledge on the image or video, even if used implicitly. In the context of single image denoisers, leading algorithms rely on sparse representations \cite{NCSR_REF,IM_DENOISE_SPARSE_3D, IM_DEMOISING_SPARSE, IM_DENOISING_IMPROVED_SPARSE}, self-similarity \cite{IM_DENOISE_SELF_SIM, BM3D_DENOISE},
and more \cite{PLOW_REF, GLOBAL_IM_DENOISE, SIAST_REF, WNNM_REF}. In addition, highly effective deep learning solutions of this problem are also available \cite{IM_DENOISE_CNN, IRCNN_REF, FFDNET_REF}. The performance of all these methods (deep-learning-based  and others) is so good that recent work investigated the possibility that we are nearing a  performance limit \cite{DENOISNG_LIMIT, DENOISNG_LIMIT2, DENOISNG_LIMIT3}.

As expected, the priors used for image denoising are very similar to the ones used in SISR. Yet migrating these denoising algorithms to serve a different
problem, such as SISR, is difficult.
Consider the NCSR algorithm \cite{NCSR_REF}, which is a state-of-the-art denoiser. It was adapted to solve other inverse problems, one of which is the SISR task.
However, the code for NCSR-SISR and NCSR-Denoising ships in two different code packages, indicating that the migration from denoising to SISR is not achieved only by a small modification to the degradation model.

The gap between VSR and video denoising is even wider. Video denoisers such as \cite{VBM3D_REF, VBM4D_REF, RNLM_REF, FAST_VIDEO_NLM, VIDEO_NO_ESTIMATION, VIDEO_DENOISE_NO_MOTION, VIDEO_DENOISE_SPARSE}
have already abandoned the explicit motion-estimation algorithms still so commonly used by VSR. This results in very efficient, and highly effective denoisers for video noise removal. In contrast, with the exception of \cite{GENERAL_NLM,3DSKR_REF,3DSKR_FOLLOWUP}, VSR algorithms are left behind, still relying on explicit optical flow estimation.

\subsection{Our Unified Super-Resolution Framework}\label{SSS:Unified}
This work's objective is to create a robust and unified framework for super resolution that can be applied to both SISR and VSR problems. Our goal is to propose a single formulation that covers both cases, leading to a single algorithm that operates on both these problems in the same manner.
The path towards achieving this goal passes through the use of the PPP/RED schemes, and this implies that we shall also rely on image/video denoisers. Our starting point is the MAP formulation in Equation (\ref{MAP_SR_cost}),
\begin{align*}
\xopt = \argmin_\x \PrimGap + R\left(\x\right).
\end{align*}
We choose to interpret this expression in two different ways. For the single image case, $\x$ and $\y$ are single images, and $R(\x)$ is a single image prior. When moving to image sequences, the same equation remains relevant, where this time $\x$ and $\y$ are assumed to be complete image sequences (i.e., two volumetric datasets), and now the blur and the subsampling operations are assumed to apply to each of the frames in the sequence independently. As for $R(\x)$, it represents a video prior, able to grade complete video solutions, by taking into account both inter-frame relations (as given in Equation (\ref{Flow_Notation})), along with intra-frame dependencies.

Now that the two problems are posed as a common energy minimization objective, the PPP and RED schemes as described in Section \ref{SS:background} are applicable and relevant to our needs. All that should be done is to replace the general degradation operator $\DegradeMat$ in Equations (\ref{x_update})-(\ref{u_update} by $\ScaleMat \BlurMat$.  More specifically, Equation (\ref{v_update}) is solved differently in the PPP and the RED schemes, yet both methods use a denoiser to solve this step, be it a single image denoiser or a video one. Equation (\ref{x_update}) is quadratic in $\x$ and can be solved using simple Linear Algebra algorithms, such as conjugate gradient and similar tools. Therefore, both schemes lead to a sequence of denoising computations, surrounded by simple algebraic operations. To summarize, Equations (\ref{x_update})-(\ref{u_update}) are relevant to both the single image and the video cases. As we move from SISR to VSR, the only difference is this: \emph{The denoiser to be applied in Equation (\ref{v_update}) should a video denoiser, operating on the complete video volume $\x^{k+1}+\u^k$ at once, thus exploiting both inter- and intra redundancies.}

We note that we could have proposed a more classical frame-by-frame MAP estimator for VSR in the spirit of the work reported in \cite{WARP_REF}. This would have been done by embarking from Equation (\ref{MAP_SR_cost}) and inserting the geometrical warp relations shown in Equation (\ref{Flow_Notation}) into the log-likelihood term.
However, the warp operators $\FlowMat_{i}$ are not known in advance, and therefore a preceding step of sub-pixel motion estimation would have been needed in order to estimate them.
As a result, such a frame-by-frame VSR framework becomes a sequence of motion estimations followed by single-frame denoising steps. Thus, the classical frame-by-frame approach cannot simply apply the existing SISR framework, but rather solves first another challenging task of motion estimation. The approach we have proposed above overcome all these difficulties by deferring the inter-frame relations into the video prior.

Algorithm \ref{unified_algorithm} formulates the unified super-resolution scheme for images and videos using PPP and RED. The algorithm follows closely Equation (\ref{ADDM_iteration}) and its adaptations discussed in sections \ref{SS:PPP} and \ref{SS:RED}.
The last two steps in the for-loop are a small modification to Equation  (\ref{ADDM_iteration}), and their goal is to improve the convergence of the scheme.
We shall elaborate more on this modification in the following subsection.

\RestyleAlgo{boxed}
\begin{algorithm2e}
 \KwIn{ $\y$ -- a LR image/video; \hspace{\textwidth} $D(\x,\sigma)$ -- image/video denoiser, cleaning an image/video $ \x $ contaminated by noise with std $ \sigma $; \hspace{\textwidth} $\sigma$ -- The noise level in $\y$  \hspace{\textwidth} $\ScaleMat$ -- The scaling operator;  \hspace{\textwidth}
 $\BlurMat$ -- The blur operator;
 \hspace{\textwidth}$ \beta$ -- parameter of confidence in the prior;
 \hspace{\textwidth}$ \rho$ -- ADMM penalty parameter;
 \hspace{\textwidth}$ \alpha$ -- ADMM penalty parameter update factor;
 \hspace{\textwidth}$ iter$ -- Number of iterations.
  \hspace{\textwidth}$ iter_{inner}$ -- Number of iterations in fixed-point method when  using RED.}
 \KwOut{a SR image/video}
\vspace{0.2cm}
\textbf{Initialization:}  $\u^0 = \veczero;$ \hspace{\textwidth} $\rho^0 = \rho; $ \hspace{\textwidth} $\x^0 = \v^0 = bicubic\_interpolation(\y);$ \hspace{\textwidth} $L = \frac{1}{\sigma^2}\left(\ScaleMat\BlurMat\right)^T\ScaleMat\BlurMat$. \\
\vspace{0.2cm}
\For{ $k = 1:1:iter$}{
\textbullet~$\x^{k+1} = \left(L+\rho^k \mathbf{I}\right)^{-1}\left(\frac{1}{\sigma^2}\left(\ScaleMat\BlurMat\right)^T\y+\rho^k\left(\v^k-\u^k\right)\right)$ \\
\textcolor{blue}{
\uIf{PPP}{
	\textbullet~$\v^{k+1} = D\left(\x^{k+1} + \u^k, \sqrt{\frac{\beta}{\rho^k}}\right)$ \\
}}
\textcolor{red}{
\Else{
// RED \\
	\textbullet~$z^0 = \v^{k}$ \\
	\For{ $j = 0:1:iter_{inner}-1$ }{
		\textbullet~$z^{j+1} = \frac{1}{\beta +  \rho^k}\left(\beta D\left(z^{j}, \sqrt{\frac{\beta}{\rho^k}}\right) +  \rho^{k}\left(\x^{k+1} + \u^k\right)\right)$ \\
	}
	\textbullet~$\v^{k+1} = z_{iter_{inner}}$ \\
}}
\textbullet~Estimate the dual gap by computing $\|\rho^{k}\left(\v^{k+1} - \v^{k}\right)\|_2^2$, and decrease $\rho^{k+1}$ if this measure constantly increases. Otherwise $\rho^{k+1} = \alpha\rho^k$ \\
\textbullet~$\u^{k+1} = \frac{\rho^k}{\rho^{k+1}}\left(\u^{k} + \x^{k+1} - \v^{k+1}\right)$ \\
 }
\vspace{0.2cm}
\KwRet{$\v^{k+1}$}
\vspace{0.2cm}
 \caption{Our proposed scheme for turning an image/video denoiser into a super-resolution solver using PPP/RED.}  \label{unified_algorithm}
\end{algorithm2e}

As expected, RED (in red) and PPP (in blue) differ only in the $\v$-update stage. Both apply a denoiser in this stage, yet PPP applies it only once, whereas RED applies it as part of a fixed-point iteration -- possibly several times.
As denoising is the most time consuming operation during the iteration, the above implies that RED is expected to be slower as $iter_{inner}$ increases.

\subsection{Convergence} \label{SSS:Convergence}
It is shown in \cite{ADMM_REF} that ADMM is guaranteed to converge under two conditions:
\begin{itemize}
  \item The two terms being minimized in Equation (\ref{MAP_cost_split}), that is, the log-likelihood term and the prior, are closed, proper and convex, and
  \item The unaugmented Lagrangian has a saddle point.
\end{itemize}
Optimality conditions are primal and dual feasibility. Primal feasibility is reached when $\x=\v$. Increasing $\rho$ will increase the penalty for $\|\x - \v\|_2^2$ and hence will guarantee primal feasibility.
On the other hand, \cite{ADMM_REF} stresses that for dual feasibility, $\rho \left(\v_{k+1}-\v_{k}\right) \rightarrow 0$ must hold.
Hence, to achieve primal feasibility $\rho$ should increase, but in a manner that prevents the increase of $\rho \left(\v_{k+1}-\v_{k}\right)$, so dual feasibility may be achieved as-well. $\rho$'s update in Algorithm \ref{unified_algorithm} is meant to increase $\rho$ as long as $\rho \left(\v_{k+1}-\v_{k}\right)$ keeps decreasing.
Since $\u$ is the \emph{scaled} Lagrange multipliers vector, a change in $\rho$ demands a rescale of $\u$. Hence, Equation (\ref{u_update}) is rescaled by the factor $\frac{\rho^k}{\rho^{k+1}}$ in the algorithm.

PPP's convergence was discussed in \cite{PPP_REF,PPP_CONVERGENCE_REF}, yet, since the prior is implicit in this scheme, convergence is not guaranteed in general.
RED, on the other hand, is known to converge under mild conditions (see \cite{RED_REF} for more details).
Indeed our experimental results in section \ref{SS:Results} show this very well.

\section{Experimental Results}\label{SS:Results}
In this section we present various experimental results\footnote{All tests were conducted on a computer running Windows 8.1, with an Intel Core i7-4500U CPU 1.80GHz and 8GB RAM installed.} that demonstrate the effectiveness of our scheme.
In our previous work \cite{OUR_SISR_REF}, we tested the PPP scheme on the SISR problem, aiming to increase the resolution of a single LR image.
The SISR problem is considered simpler, and less time consuming than VSR, and hence tuning our framework is made easier in this case.
We used the NCSR \cite{NCSR_REF} algorithm, both as a denoiser in Algorithm \ref{unified_algorithm} and as a super-resolution algorithm to compare with.
The proposed approach proved successful and the experiments are detailed in \cite{OUR_SISR_REF}.
The transition to VSR leads to a counter-intuitive paradigm that shows how the VSR problem can be handled \emph{without relying on an accurate and explicit motion-estimation algorithm}. A careful test of this core idea is the focus of this work, and is therefore detailed in the remainder of this section.

We use the VBM3D algorithm \cite{VBM3D_REF} as a video denoiser both in PPP and RED, as it provides state-of-the-art denoising results; it is motion-estimation free, and hence very fast and efficient.
By doing so we achieve a VSR algorithm which is motion-estimation free, and benefits from the efficiency of the chosen denoiser.
In this section, we show that the resulting algorithm is indeed more efficient than existing VSR algorithms, without compromising quality of the final result.
We use the same tuned parameters as in the SISR case \cite{OUR_SISR_REF} but with a small increase in the number of iterations:
\begin{align*}
\rho_0 = 0.0001, \quad \beta = 0.2048, \quad \alpha = 1.2, \quad ~iter=40,
\end{align*}
where $\rho_0$ is the initial penalty parameter, $\beta$ is the confidence in the prior, $\alpha$ is the penalty parameter step, meaning each iteration $\rho$ will be multiplied by $\alpha$ (unless the dual gap increases) and $iter$ is the number of iterations.
For RED, the number of fixed-point iterations, $iter_{inner}$, should be set as well. RED-1 represents a setting where $iter_{inner}=1$ and similarly, for RED-2 $iter_{inner}=2$.
We compare our framework on several scenarios:
\begin{enumerate}
  \item Single frame super resolution from multiple frames of global translations.
  \item Single frame super resolution from real videos.
  \item Super-resolved video from real videos.
\end{enumerate}
The scenarios and their corresponding experiments are depicted in the following subsections. 
\subsection{Single frame super resolution from multiple frames of global translations}
In this test our input is a group of multiple frames, which are all a global translation of the first frame. The goal of this synthetic experiment is to validate that the proposed scheme leads to  a truly super-resolved outcome in a controlled case, and verify that it extracts most of the aliasing for producing this result. 

The translation is chosen randomly for each frame, up to 5 pixels in each axis.
We generate 30 frames, blurred with a Gaussian kernel of s.t.d. 1 and size $3\times3$, then down-sampled by factor 2 and  contaminated with a white Gaussian noise of s.t.d. $\sqrt2$.
On these LR frames we run the shift-and-add algorithm reported in \cite{FAST_ROBUST_MFSR} (referred to hereafter as TIP04), which suggests a fast and robust algorithm for recovering a high-resolution image from the group of LR global translationed, blurred and noisy versions of it. TIP04 minimizes an $L_1$ energy term and uses a Biliteral-TV  regularization. A fast implementation is suggested for pure translations, and this is the one we use. 

\begin{figure}[!htbp]
	\centering
	\subfloat[Original]{\includegraphics[scale=0.30]{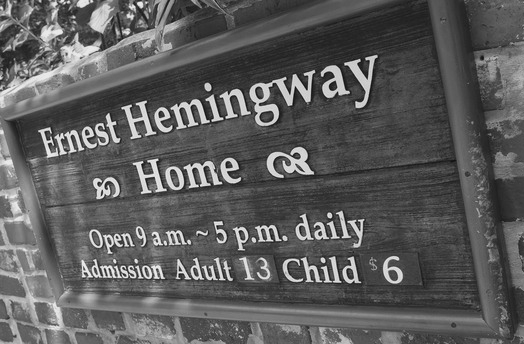}}
	\centering
	\hfil
	\subfloat[LR]{\includegraphics[scale=0.30]{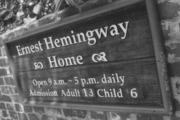}}
	\centering
	\hfil
	\subfloat[Bicubic (21.07 dB)]{\includegraphics[scale=0.30]{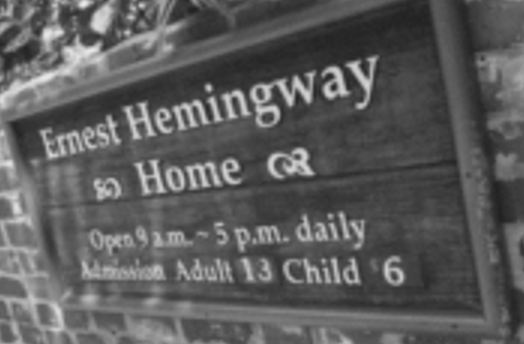}}
	\quad
	\includegraphics[scale=1]{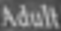}
	\centering
	\hfil
	\subfloat[TIP04 (25.03 dB)]{\includegraphics[scale=0.30]{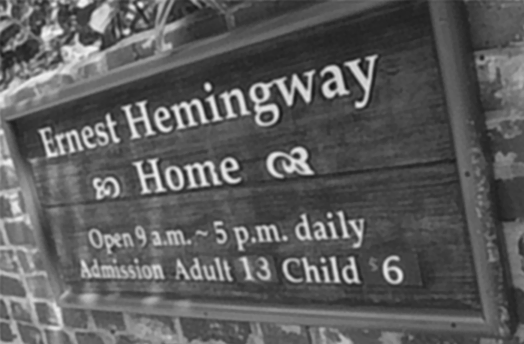}}
	\quad
	\includegraphics[scale=1]{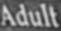}
	\centering
	\hfil
	\subfloat[RED-2 (26.48 dB)]{\includegraphics[scale=0.30]{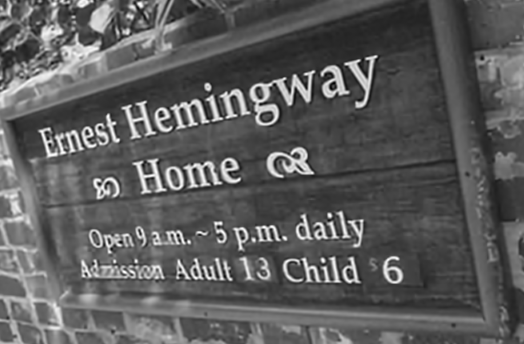}}
	\quad
	\includegraphics[scale=1]{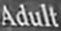}
	\centering
	\caption{ TIP04 and RED results on a real image and a zoom-in on the word \emph{Adult}.}
	\label{TIP04_allias}
\end{figure}

We compare the results to RED-2, by treating the whole set of frames jointly, reconstructing a whole SR video, and then taking only the first from the outcome.
For both competing methods we compute the Peak Signal to Noise Ratio (PSNR\footnote{PSNR$\left(X,Y\right)=10\log_{10}\left(255^2/\frac{1}{P}\|X-Y\|_2^2\right)$), where $P$ is the size of the image.}) of the first frame (without its borders).
Notice that our algorithm is unaware of the fact that the input video is just a translation of the first frame, whereas TIP04 relies on this  knowledge explicitly.
TIP04 shows exceptionally good results for images with large and smooth edges, yet when the details became smaller and sharper, TIP04 encounters difficulties, and RED-2 outperforms it, as can be seen in Figure \ref{MultiFrameTip}.
Figure \ref{TIP04_allias} presents the results of a second and similar experiment on a real world image, down-sampled with factor 3.
Observe the aliasing in the word "Adult" in the bicubic restoration (mainly in 'l' and 't'), which has no trace in super-resolved results of the two competing methods. 

These two experiments we have just described are characterized by exhibiting a simple and global motion, for which classic super-resolution methods, such as TIP04, are very effective.
In such scenarios, a near perfect super-resolved outcome can be expected, recovering small details immersed in strong aliasing effects.
The goal in these tests was to verify that the proposed algorithms maintain this super-resolution capability.
The results indicate that, indeed, our methods successfully resolve higher resolution images, being competitive with state-of-the-art methods that are explicitly designed for this regime.
We now turn to more challenging experiments with more complex video content, for which classic methods are expected to fail.

\begin{table}[ht]
\begin{tabular}{ccc}
\begin{sideways}Original\end{sideways} &
\subfloat{\includegraphics[scale=1]{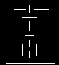}} &
\subfloat{\includegraphics[scale=1]{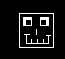}} \\ \newline
\begin{sideways}TIP04\end{sideways} &
\subfloat[(17.12 dB)]{\includegraphics[scale=1]{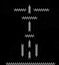}} &
\subfloat[(22.15 dB)]{\includegraphics[scale=1]{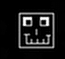}} \\ \newline
\begin{sideways}RED-2\end{sideways} &
\subfloat[(24.96 dB)]{\includegraphics[scale=1]{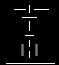}} &
\subfloat[(26.66 dB)]{\includegraphics[scale=1]{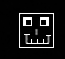}} \\ \newline
\end{tabular}
\begin{tabular}{cc}
\begin{sideways}Original\end{sideways} &
\subfloat{\includegraphics[scale=0.8]{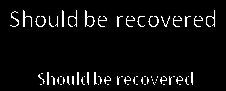}}  \\ \newline
\begin{sideways}TIP04\end{sideways} &
\subfloat[(19.84 dB)]{\includegraphics[scale=0.8]{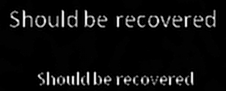}}  \\ \newline
\begin{sideways}RED-2\end{sideways} &
\subfloat[(25.08 dB)]{\includegraphics[scale=0.8]{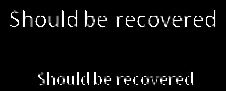}} \\ \newline
\end{tabular}
    \captionof{figure}{Three image reconstructions from a group of images with pure translations between them.}
\label{MultiFrameTip}
\end{table}

\subsection{Single frame super resolution from real videos}
The recently published DeepSR \cite{DEEPSR_REF} is a state-of-the-art algorithm that aims to solve a slightly different problem than the classic VSR: Given the whole LR video, instead of restoring the entire sequence, DeepSR estimates only the mid-frame.  It does so in two steps: First, several SR estimates for the mid-frame are generated from the LR video using different motion-estimations. The second stage is to merge HR details into a single frame by feeding all the above estimates to a trained CNN.
The software package for DeepSR is available online \cite{DEEP_SR_LINK}, along with the dataset it was tested on.
The code includes the pre-configured hyper parameters and a pre-trained CNN model.
The provided package assumes a Gaussian blur of s.t.d. within the range of 1.2 to 2.4. The LR videos are created by (i) blurring each HR frame with a $ 7 \times7 $ Gaussian kernel of s.t.d. 1.5, followed by (ii) down scaling by a factor 4 in each axis, and (iii) adding a Gaussian noise with $ \sigma = 1 $ to the outcome.

We also find it interesting to compare our algorithm also to a trivial extension of SISR to VSR, hence we tested IRCNN \cite{IRCNN_REF}, which is a state-of-the-art SISR method, on the same data set (applying it frame by frame). IRCNN uses a trained CNN that learned denoising priors to form a new denoiser that aids in solving inverse problems in a manner similar to PPP or RED.

\begin{figure}[!htbp]
	\centering
	\subfloat[DeepSR (31.133 dB)]{\includegraphics[scale=0.8]{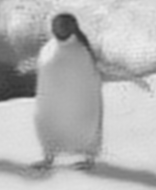}}
	\centering
	\hfil
	\subfloat[IRCNN (33.0867 dB)]{\includegraphics[scale=0.8]{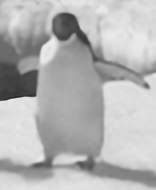}}
	\centering
	\hfil
	\subfloat[PPP (37.137 dB)]{\includegraphics[scale=0.8]{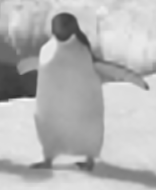}}
	\centering
	\hfil
	\subfloat[RED-2 (38.0377 dB)]{\includegraphics[scale=0.8]{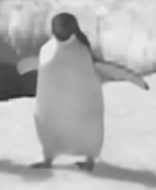}}
	\centering
	\hfil
	\subfloat[Original]{\includegraphics[scale=0.8]{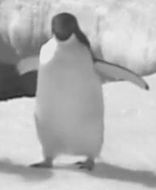}}
	\centering
	\caption{A zoomed-in area of the mid-frame of the \textsf{Penguin} sequence, along with the corresponding PSNR.}
	\label{Penguin_SR_result_penguin}
\end{figure}

\begin{figure}[!htbp]
	\centering
	\subfloat[DeepSR]{\includegraphics[scale=0.8]{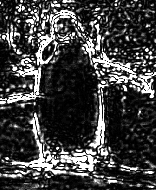}}
	\centering
	\hfil
	\subfloat[IRCNN]{\includegraphics[scale=0.8]{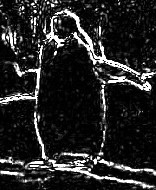}}
	\centering
	\hfil
	\subfloat[PPP]{\includegraphics[scale=0.8]{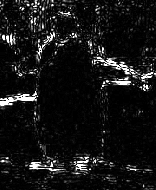}}
	\centering
	\hfil
	\subfloat[RED-2]{\includegraphics[scale=0.8]{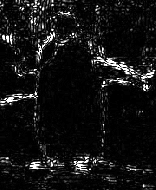}}
	\centering
	\caption{Squared error of the areas presented in Figure \ref{Penguin_SR_result_penguin}}
	\label{Penguin_SR_result_penguin_err}
\end{figure}

Table \ref{DEEPSR_psnr_comp} compares our PPP and RED schemes to the bicubic interpolation and IRCNN, where the PSNR is averaged over the entire sequence of frames in each video. As can be seen, RED outperforms PPP, and both lead to better reconstructions than the bicubic and IRCNN. Table \ref{DEEPSR_psnr_midframe_comp} compares the proposed algorithms to DeepSR, where we measure the PSNR on the luminance channel of the mid-frame of each video. On average, RED is leading this table, the second best approach being the PPP, and both results lead to better reconstruction than DeepSR and IRCNN.
Figure \ref{Penguin_SR_result_penguin} compares visually cropped regions that are extracted from the recovered mid-frames of the \textsf{Penguin}  video. As can be seen, DeepSR suffers from artifacts around fast moving objects such as the Penguin's wings. Figure \ref{Penguin_SR_result_penguin_err} shows the squared error for the same images.

\begin{table}[ht]
    \begin{tabular}{ | l | c | c | c | c | c |}
    \hline    
    \textbf{Video / Alg.} & \textbf{Bicubic} & \textbf{IRCNN} & \textbf{PPP} & \textbf{RED-2} \\ \hline
    \hline
	\textsf{Calendar}	& 18.74  & 20.77	 & 21.45 	& \textbf{21.53} \\ \hline
	\textsf{City}	& 23.81  & 25.12  & \textbf{26.34}	 	& 26.30 \\ \hline
	\textsf{Foliage}	& 21.54	 & 23.58  & \textbf{25.01}	  & 24.99 \\ \hline
	\textsf{Penguin}	& 28.80	 & 33.78  & 34.55	 	& \textbf{35.54} \\ \hline
	\textsf{Temple}	& 24.30	& 27.25  & 29.81		& \textbf{29.98} \\ \hline
	\textsf{Walk}	& 23.01	& 26.36  & 28.49		& \textbf{28.57} \\ \hline
	 \hline
	\textbf{Average} & 23.37 & 26.14  & 27.60  & \textbf{27.82} \\ \hline
    \end{tabular}
	
    \captionof{table}{PSNR [dB] comparison (averaged over the entire sequence) between the bicubic interpolation, IRCNN and our algorithms. The best results are highlighted.} \label{DEEPSR_psnr_comp}
\end{table}

\begin{table}[ht]
    \begin{tabular}{ | l || c | c | c | c | }
    \hline
    \textbf{Video / Alg.} & \textbf{IRCNN} & \textbf{DeepSR}  & \textbf{PPP} & \textbf{RED-2}\\ \hline
    \hline
	\textsf{Calendar} &20.76 	& 21.53	&  21.53	 &  \textbf{21.62}	\\ \hline
	\textsf{City}	& 24.54 & \textbf{25.83}	& 25.52 	 & 25.50 \\ \hline
	\textsf{Foliage} & 23.48	&  \textbf{24.95}	& 24.88	 & 24.88 \\ \hline
	\textsf{Penguin} & 33.88 	& 32.10	& 34.56	 &   \textbf{35.57} \\ \hline
	\textsf{Temple}& 27.49	& 30.60	&  30.77	 &  \textbf{30.81}	\\ \hline
	\textsf{Walk} & 26.42	& 26.46	& 28.54	 &  \textbf{28.59} \\ \hline
	 \hline
	\textbf{Average} & 26.01 & 26.91 & 27.63  & \textbf{27.83} \\ \hline
    \end{tabular}
    \captionof{table}{PSNR [dB] comparison between our algorithms, IRCNN and DeepSR (PSNR computed only on the midframe of our restoration). The best results are highlighted.} \label{DEEPSR_psnr_midframe_comp}
\end{table}

\begin{table}[ht]
    \begin{tabular}{ | l || c | c | c |c| }
    \hline
    \textbf{Video / Alg.} & \textbf{DeepSR} & \textbf{IRCNN} & \textbf{PPP}  & \textbf{RED-2}\\ \hline \hline
	\textsf{Calendar}	& 3983 &	4462 &  \textbf{2421}	 &4563 \\ \hline
	\textsf{City}	& 3929	& 4419&  \textbf{2367}	 & 3816\\ \hline
	\textsf{Foliage}	& 3372	& 3695 &  \textbf{1965}	& 3094 \\ \hline
	\textsf{Penguin}	& 11574 & 11113	& \textbf{ 5321}	& 8360 \\ \hline
	\textsf{Temple}	& 12031 & 10465	&  \textbf{5874}	& 9116 \\ \hline
	\textsf{Walk}	& 3359 & 3687	&  \textbf{1951}	 & 3049 \\ \hline
	 \hline
	\textbf{Average} & 6375 & 6307 & \textbf{3317} & 5333 \\ \hline
    \end{tabular}
    \captionof{table}{Duration of each algorithm measured [sec]. DeepSR only reconstructs the midframe while all the others reconstruct 31 frames. The best results are highlighted.} \label{DEEPSR_duration_comp}
\end{table}

Table \ref{DEEPSR_duration_comp} displays the time consumption of each algorithm per video. It is important to stress that we restore the entire sequence of frames in the reported time, whereas DeepSR reconstructs only the mid-frame. One can see that the PPP scheme restores the whole video in about half of the time that it takes for DeepSR to restore a single frame. As for RED-2, it is slower, but still faster then DeepSR. On average IRCNN is slower than applying RED-2.

\begin{figure}[!htbp]
	\centering
	\subfloat[Input]{\includegraphics[scale=1]{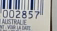}}
	\centering
	\hfil
	\subfloat[DeepSR]{\includegraphics[scale=1]{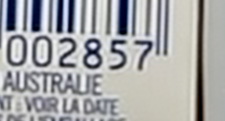}}
	\centering
	\hfil
	\subfloat[RED-2]{\includegraphics[scale=1]{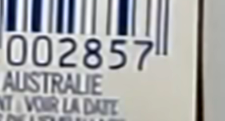}}
	\centering
	\caption{Midframe of a real low resolution video constructed by DeepSR and RED-2.}
	\label{REAL_DeepSR_sr}
\end{figure}

Figure \ref{REAL_DeepSR_sr} shows a comparison between DeepSR and RED-2 on the \textsf{Barcode} sequence, a real low-resolution video with no ground truth. One may notice that DeepSR suffers from halos, which do not appear in RED-2's output.

\subsection{Super-resolved video from real videos}
3DSKR \cite{3DSKR_REF} is a VSR algorithm free of explicit sub-pixel motion-estimation, and thereby capable of processing videos with complex motion. Specifically, the 3DSKR is composed of two-stages; The first is a super-resolution process (this step is formulated as a weighted Least-Squares problem that captures the local motion trajectories) that ignores the blurring kernel, while the second is a deblurring step (can be thought of as a post-processing operation) that takes into account the blur kernel.
Using the code supplied by the authors \cite{3DSKR_LINK}, we perform a comparison of our methods and 3DSKR on a standard dataset that contains several grayscale videos: \textsf{Coastguard}, \textsf{Bicycle}, \textsf{MissAmerica}, \textsf{Tennis} and \textsf{Salesman}.
The 3DSKR package does not include the de-blurring phase. Following previous work \cite{MOTION_ESTIMATION_REF1} and as suggested in the supplied code \cite{3DSKR_LINK}, we further improve the performance of this method by adding the state-of-the-art BM3D deblurring \cite{BM3D_DEB_REF} algorithm as a post processing step.
The parameters of the deblurring are tuned to achieve the highest PSNR. For each video, we apply the same blur kernel as in 3DSKR's demo (average blur of size 3), same decimation (factor 3 in each axis) and add the same Gaussian noise with $ \sigma = 2 $.
Table \ref{3DSKR_psnr_comp} presents the PSNR score of each recovered video (averaged over the frames), estimated by the bicubic interpolation, 3DSKR, and the proposed algorithms.
Since 3DSKR does not restore the borders of the video nor the first frame, we did not take into consideration these parts in the PSNR computation.
The results in Table \ref{3DSKR_psnr_comp} suggest that RED-2 is the best performing algorithm, followed by the PPP, and these two outperform the 3DSKR and bicubic methods.
\begin{table}[t]
	\setlength{\tabcolsep}{5pt}				
	\renewcommand{\arraystretch}{1}
	\renewcommand{\tabcolsep}{4pt}
	\begin{center}
	{\begin{tabular}{ | l || c | c |  c | c | c | c|}
    \hline
    \textbf{\pbox{30cm}{Video / Alg.}} & \textbf{Bicubic} & \textbf{\pbox{30cm}{3DSKR\\+Deblur}} & \textbf{PPP} & \textbf{RED-1} & \textbf{RED-2} \\ \hline
        \hline
	\textsf{Coastguard} & 23.77& 24.75& 25.16& 25.16& \textbf{25.27}\\ \hline
	\textsf{Bicycle} & 21.62&  24.32& 28.79& 28.80& \textbf{29.15}\\ \hline
	\textsf{Foreman} & 27.97& 31.15& 33.43& 33.43 & \textbf{33.53}\\ \hline
	\textsf{Salesman} & 24.18& 25.96& 26.64& 26.638& \textbf{26.77}\\ \hline
	\textsf{MissAmerica} & 32.10& 35.55& 36.85& 36.85& \textbf{37.20}\\ \hline
	\textsf{Tennis} & 21.91& 22.78& 23.14& 23.14 & \textbf{23.16}\\ \hline
	 \hline
	\textbf{Average} & 25.26& 27.42& 29.00& 29.00& \textbf{29.18}\\ \hline
    \end{tabular}}
    \captionof{table}{PSNR [dB] comparison between the bicubic, 3DSKR+deblurring, and our algorithms (PPP and RED) (RED-X is RED with X inner iterations). The PSNR is computed only on the areas restored by 3DSKR. The best results are highlighted.}
    \label{3DSKR_psnr_comp}
\end{center}
\end{table}

Table \ref{3DSKR_duration_comp} depicts	the runtime for the different videos and algorithms, indicating that a massive boost in runtime is also achieved. Specifically, the PPP is 30 to 50 times faster than 3DSKR (average factor of 42); RED is slightly behind, with a gain in speed-up that is approximately 23.
Note that PPP is roughly 2 times faster than RED-2, as the later applies the denoiser twice in each iteration, while PPP calls the denoiser only once. This aligns with the observation that the denoising operation is the most time consuming step in our algorithms. At this point we should stress that the Fixed-Point algorithm, suggested in RED \cite{RED_REF}, might be the key to obtain a much faster process. We defer this for a future work.

\begin{table}[t]
		\setlength{\tabcolsep}{5pt}				
		\renewcommand{\arraystretch}{1}
		\renewcommand{\tabcolsep}{4pt}
\begin{center}
    {\begin{tabular}{ | l || c | c | c | c | c|}
    \hline
    \textbf{\pbox{30cm}{Video / Alg.}} & \textbf{3DSKR} & \textbf{\pbox{30cm}{3DSKR\\+Deblur}} & \textbf{PPP} & \textbf{RED-1} & \textbf{RED-2}\\ \hline
    \hline
	\textsf{Coastgaurd}  &	3268  &	3282 &	92 &	\textbf{91} &	156 \\ \hline
	\textsf{Bicycle}	 & 73485	 & 73748	 & 1835 & \textbf{1780}	 & 3170\\ \hline
	\textsf{Foreman}	 & 16590	 & 16649	 & \textbf{373} & 	381 & 	686\\ \hline
	\textsf{Salesman}	 & 15020	 & 15081	 & \textbf{343}	 & 351 & 662 \\ \hline
	\textsf{MissAmerica}	 & 16463	 & 16524	 & 385 & \textbf{377} & 659\\ \hline
	\textsf{Tennis}	 & 13108	 & 13159	 & \textbf{290}	 & 300 & 560 \\ \hline
	 \hline
	\textbf{Average}	& 22989 	& 23074	& 553	& \textbf{547} & 982 \\ \hline
    \end{tabular}}
    \captionof{table}{Runtime [sec] of 3DSKR, 3DSKR+Deblurring, and our algorithms (RED-X is RED with X inner iterations). The best results are highlighted.}
    \label{3DSKR_duration_comp}
\end{center}
\end{table}

\begin{figure}[!htbp]
\centering
  \subfloat[Original]{\includegraphics[scale=1.2]{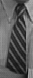}} 
   \hfil
  \centering
  \subfloat[Bicubic  \newline (22.72 dB)]{\includegraphics[scale=1.2]{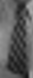}} 
   \hfil
  \centering
  \subfloat[3DSKR+BM3D \newline(23.98 dB)]{\includegraphics[scale=1.2]{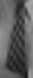}} 
   \hfil
  \centering
  
  \subfloat[PPP \newline(25.65 dB)]{\includegraphics[scale=1.2]{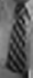}}  
   \hfil
  \centering
  \subfloat[RED-1 \newline(25.62 dB)]{\includegraphics[scale=1.2]{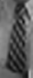}}  
   \hfil
  \centering
  \subfloat[RED-2 \newline (26.00 dB)]{\includegraphics[scale=1.2]{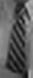}} 
\caption{Zoomed in versions of the tie region (and the corresponding PSNR), extracted from \textsf{Salesman}.}
\label{salesman_25_SR_results_tie}
\end{figure}

\begin{figure}[!htbp]
\centering
  \subfloat[Original]{\includegraphics[scale=0.9]{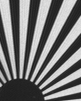}}
  \hfil
  \centering
  \subfloat[Bicubic (14.40 dB)]{\includegraphics[scale=0.9]{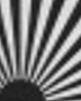}}
  \hfil
  \centering
  \subfloat[3DSKR+BM3D \newline (17.88 dB)]{\includegraphics[scale=0.9]{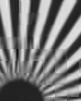}}
  \hfil
  \centering
  \subfloat[PPP (22.74 dB)]{\includegraphics[scale=0.9]{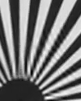}}
  \hfil
  \centering
  \subfloat[RED-1 (22.69 dB)]{\includegraphics[scale=0.9]{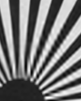}}
  \hfil
  \centering
  \subfloat[RED-2 (26.22 dB)]{\includegraphics[scale=0.9]{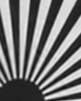}}
\centering
\caption{Zoomed in versions of the striped region (and the corresponding PSNR), extracted from \textsf{Bicycle}. }
\label{bicycle_10_SR_results_stripe}
\end{figure}

Returning to the outcome visual quality, one might falsely deduce that the small increase in PSNR between PPP and RED-2 does not justify the increase in time consumption. Yet, a closer look at the results shows that RED-2 performs much better. Figure \ref{salesman_25_SR_results_tie} shows how RED-2 restores the pattern of the tie, while all the other algorithms confuse it with a tile-like pattern due to aliasing. Figure \ref{bicycle_10_SR_results_stripe} shows that RED-2 suffers less from ``pixelized'' edges across the stripes. 

\begin{figure}[!htbp]
\includegraphics[scale=0.5]{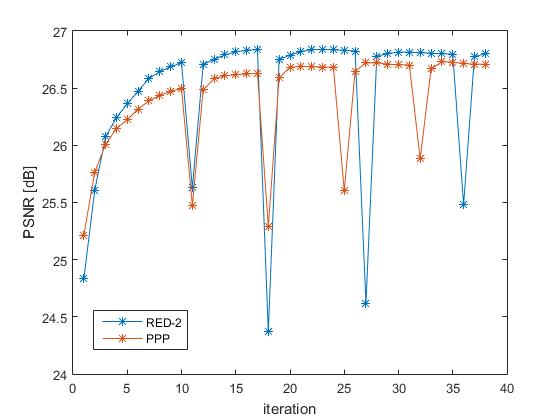}
\caption{PSNR during iteration of PPP and RED-2 on the \textsf{Salesman} sequence.}
\label{salesman_convergence}
\end{figure}

Figure \ref{salesman_convergence} displays the PSNR during the iterations of RED-2 and PPP on the \textsf{Salesman} sequence. The sharp drop in PSNR occurs when $\rho$ is decreased to ensure dual feasibility.
One can see that PPP converges more slowly, and is highly dependent on the $\rho$-update (the PSNR is flattened until $\rho$ is decreased). RED, on the other hand, converges faster, and could have stopped the iteration earlier with almost the same PSNR score. 


\begin{figure}[!htbp]
\centering
  \subfloat[Original]{\includegraphics[scale=1]{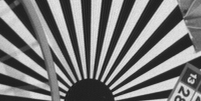}}
  \hfil
  \centering
  \subfloat[LR]{\includegraphics[scale=1]{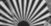}}
  \hfil
  \centering
  \subfloat[Bicubic (14.13 dB)]{\includegraphics[scale=1]{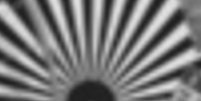}}
  \hfil
  \centering
  \subfloat[SPMC (16.66 dB)]{\includegraphics[scale=1]{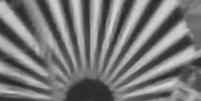}}
  \hfil
  \centering
  \subfloat[RED-2 (21.89 dB)]{\includegraphics[scale=1]{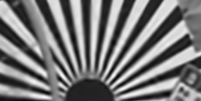}}
\centering
\caption{Zoomed in patch of the \textsf{Bicycle} video, frame 10 (and the corresponding PSNR) for Bicubic, SPMC and RED-2. These results refer to the average blur case.}
\label{bicycle_spmc_compare}
\end{figure}

\begin{figure}[!htbp]
\centering
  \subfloat[Original]{\includegraphics[scale=0.7]{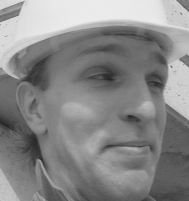}}
  \hfil
  \centering
  \subfloat[LR]{\includegraphics[scale=0.7]{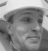}}
  \hfil
  \centering
  \subfloat[Bicubic (28.31 dB)]{\includegraphics[scale=0.7]{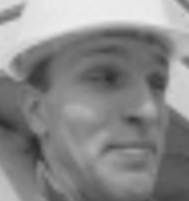}}
  \hfil
  \centering
  \subfloat[SPMC (28.94 dB)]{\includegraphics[scale=0.7]{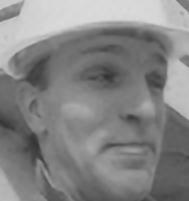}}
  \hfil
  \centering
  \subfloat[RED-2 (34.72 dB)]{\includegraphics[scale=0.7]{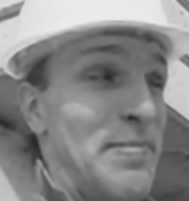}}
\centering
\caption{Zoomed in patch of the \textsf{Foreman} video, frame 5, (and the corresponding PSNR) for Bicubic, SPMC and RED-2. These results refer to the Gaussian blur case.}
\label{foreman_spmc_compare}
\end{figure}

Our last reported experiment offers a comparison with a method called SPMC \cite{SPMC_REF}. SPMC reconstructs the video frame-by-frame using a batch of frames around the reconstructed one. It is composed of several stages: motion-estimation, followed by sub-pixel motion compensation, and finally the aligned frames are given as an input to a detail-fusion CNN. The code is available online \cite{SPMC_LINK}, yet supports only scale factors 2 and 4. Therefore we conducted the same test as with 3DSKR, but with a scale factor 4.
The degradation includes an average blur and a Gaussian blur kernel of size 7x7 and s.t.d. 1.5, and an additive noise level set to $\sigma=1$. Table \ref{SPMC_psnr_comp_ave} presents the resulting PSNR when using average blur and Table \ref{SPMC_psnr_comp_gauss} for the Gaussian one.
Following Tables \ref{SPMC_psnr_comp_ave} and \ref{SPMC_psnr_comp_gauss}, SPMC yields disappointing PSNR results, but it seems that this is mainly due to a change in grayscale level in SPMC's output as can be seen in Figures \ref{bicycle_spmc_compare} and \ref{foreman_spmc_compare}.
A visual comparison shows that SPMC performs very well for the Gaussian blur, while RED-2 is roughly of the same quality. For the average blur, RED-2 seems to outperform SPMC. 

\begin{table}[ht]
	\begin{center}
	{\begin{tabular}{ | l || c | c |  c | }
    \hline
    \textbf{\pbox{30cm}{Video / Alg.}} & \textbf{Bicubic}  & \textbf{SPMC} & \textbf{RED-2} \\ \hline
        \hline
	\textsf{Coastguard} & 22.63& 22.97& \textbf{24.34}\\ \hline
	\textsf{Bicycle} & 18.88&  20.38& \textbf{24.60}\\ \hline
	\textsf{Foreman} & 24.42& 24.11& \textbf{30.56}\\ \hline
	\textsf{Salesman} & 23.06& 23.37& \textbf{25.79}\\ \hline
	\textsf{MissAmerica} & 29.53& 27.16& \textbf{34.57}\\ \hline
	\textsf{Tennis} & 21.38& 21.88& \textbf{22.76}\\ \hline
	 \hline
	\textbf{Average} & 23.31& 23.31& \textbf{27.10}\\ \hline
    \end{tabular}}
    \captionof{table}{PSNR [dB] comparison between the bicubic, SPMC, and RED-2 on \textbf{average} blur and scale=4. The best results are highlighted.}
    \label{SPMC_psnr_comp_ave}
\end{center}
\end{table}

\begin{table}[ht]
	\begin{center}
	{\begin{tabular}{ | l || c | c |  c | }
    \hline
    \textbf{\pbox{30cm}{Video / Alg.}} & \textbf{Bicubic}  & \textbf{SPMC} & \textbf{RED-2} \\ \hline
        \hline
	\textsf{Coastguard} & 22.85& 23.72& \textbf{24.48}\\ \hline
	\textsf{Bicycle} & 19.00&  23.46& \textbf{25.18}\\ \hline
	\textsf{Foreman} & 24.70& 25.85& \textbf{30.48}\\ \hline
	\textsf{Salesman} & 23.19& 24.25& \textbf{25.63}\\ \hline
	\textsf{MissAmerica} & 29.68& 27.80& \textbf{34.90}\\ \hline
	\textsf{Tennis} & 21.38& \textbf{23.02}& 22.76\\ \hline
	 \hline
	\textbf{Average} & 23.47& 24.68& \textbf{27.24}\\ \hline
    \end{tabular}}
    \captionof{table}{PSNR [dB] comparison between the bicubic, SPMC, and RED-2 on \textbf{Gaussian} blur and scale=4. The best results are highlighted.}
    \label{SPMC_psnr_comp_gauss}
\end{center}
\end{table}

\FloatBarrier

\section{Conclusion}\label{SS:Conclusion}
In this work we have presented a simple unified scheme for integrating denoisers into both single-frame and video super-resolution, relying on the PPP and RED frameworks.
The integration is done by using the denoiser as a black-box tool, applied within the iterative recovery process.
The algorithm's parameters were first tuned for SISR, the easier problem, and then used for VSR without a change.
We compared our proposed schemes to super-resolution algorithms for SISR, Multi-Frame Super-Resolution and VSR, achieving in all cases state-of-the-art results.
More specifically, using an existing and efficient video denoiser (VBM3D \cite{VBM3D_REF}) we have created a robust and powerful VSR algorithm that does not depend on good locality or explicit motion-estimation.

Future work can adopt other optimization methods, suggested in \cite{RED_REF}, such as the fixed point strategy. This may lead to a further improvement in time consumption. 
Another promising direction is modifying the proposed scheme to work on batches or even sequentially and causally rather than the entire sequence, reducing the memory requirements.
Another possibility would be to further improve the results by plugging better denoising algorithms such as the VBM4D \cite{VBM4D_REF} or others.
More work should be done regarding the choice of the parameters in the ADMM. Currently, the conditions to increase or decrease $\rho$ were set empirically according to the term $\rho\left(\v_{k}-\v_{k+1}\right)$. Other approaches are suggested in \cite{ADMM_REF}, but thresholds with better mathematical reasoning are yet to be found.

\section*{ACKNOWLEDGMENTS}
The research leading to these results has received funding
from the European Research Council under European
Union’s Seventh Framework Program, ERC Grant agreement
no. 320649, and from the Israel Science Foundation (ISF)
grant number 1770/14.  Y. Romano would like to thank the Zuckerman Institute, ISEF foundation and Viterbi fellowship from the Technion for supporting this research.

% or


\ifCLASSOPTIONcaptionsoff
  \newpage
\fi



%\bibliographystyle{IEEEtran}
\bibliographystyle{IEEEtran}
\bibliography{refs}



%\vfill


\end{document}